\documentclass[aps,prev,twocolumn,preprintnumbers,floatfix,nofootinbib]{revtex4-1}
\pdfoutput=1
\usepackage{amssymb,graphicx}
\usepackage{epstopdf}
\usepackage{mathrsfs}
\usepackage{amssymb}
\usepackage{verbatim}
\usepackage{xcolor}
\usepackage{multirow}
\usepackage{amsmath}
\usepackage{braket}
\usepackage{subfig}
\usepackage{slashed} 
\usepackage{float}
\usepackage[normalem]{ulem}
\usepackage{sidecap}
\usepackage{hyperref}
\usepackage{cancel}
\usepackage{enumerate}
\hypersetup{pdfstartview=FitV,colorlinks=true,linkcolor=blue,citecolor=red,filecolor=black,urlcolor=blue}

\usepackage[T1]{fontenc}
\usepackage[utf8]{inputenc}

\def\stacksymbols #1#2#3#4{\def\theguybelow{#2}
    \def\vp{\lower#3pt}
    \def\sp{\baselineskip0pt\lineskip#4pt}
    \mathrel{\mathpalette\intermediary#1}}

\def\intermediary#1#2{\vp\vbox{\sp
     \everycr={}\tabskip0pt
     \halign{$\mathsurround0pt#1\hfil##\hfil$\crcr#2\crcr
              \theguybelow\crcr}}}

\newcommand{\lslashslash}{%

  \raisebox{0.8ex}{%
    \scalebox{.7}{%
      \rotatebox[origin=c]{18}{$-$}%
    }%
  }%
}
\newcommand{\lslash}{%
  {%
   \vphantom{d}%
   \ooalign{\kern-.1em\smash{\lslashslash}\hidewidth\cr${\rm l}$\cr}%
   \kern.05em
  }%
}

\newcommand*\diff{\mathop{}\!\mathrm{d}}

\newcommand{\beq}{\begin{equation}}
\newcommand{\eeq}{\end{equation}}
\newcommand{\bea}{\begin{eqnarray}}
\newcommand{\eea}{\end{eqnarray}}

\newcommand{\gsim}{\lower.7ex\hbox{$\;\stackrel{\textstyle>}{\sim}\;$}}
\newcommand{\lsim}{\lower.7ex\hbox{$\;\stackrel{\textstyle<}{\sim}\;$}}

\newcommand{\trh}{T_{\mathrm{RH}}}
\newcommand{\tmax}{T_{\mathrm{max}}}

\def\rhorh{\rho_{\rm RH}}
\def\arh{a_{\rm RH}}
\def\ae{a_{\rm end}}
\def\rhoe{\rho_{\rm end}}
\def\rhomax{\rho_{\rm max}}
\def\amax{a_{\rm max}}

\def\be{\begin{equation}}
\def\ee{\end{equation}}
\def\bea{\begin{eqnarray}}
\def\eea{\end{eqnarray}}

\def\m{\mu}
\def\n{\nu}

\def\sp{\;\;\;,\;\;\;}

\def\lsim{\raise0.3ex\hbox{$\;<$\kern-0.75em\raise-1.1ex\hbox{$\sim\;$}}}
\def\gsim{\raise0.3ex\hbox{$\;>$\kern-0.75em\raise-1.1ex\hbox{$\sim\;$}}}

\def\inbar{\,\vrule height1.5ex width.4pt depth0pt}

\def\IC{\relax\hbox{$\inbar\kern-.3em{\rm C}$}}
\def\IQ{\relax\hbox{$\inbar\kern-.3em{\rm Q}$}}
\def\IR{\relax{\rm I\kern-.18em R}}
 \font\cmss=cmss10 \font\cmsss=cmss10 at 7pt
\def\IZ{\relax\ifmmode\mathchoice
 {\hbox{\cmss Z\kern-.4em Z}}{\hbox{\cmss Z\kern-.4em Z}}
 {\lower.9pt\hbox{\cmsss Z\kern-.4em Z}}
 {\lower1.2pt\hbox{\cmsss Z\kern-.4em Z}}\else{\cmss Z\kern-.4em Z}\fi}

\def\tmax{T_{\rm max}}
\def\trh{T_{\rm RH}}

\def\comment#1{}
\def\to{\rightarrow}

\def\u1x{U(1)_X}
\newcommand{\nc}{\newcommand}
\nc{\LL}{L}
\nc{\vv}{\tilde{v}}
\nc{\ccdot}{\!\cdot\!}
\nc{\gsm}{G_{SM}}
\nc{\vfive}{\mathbf{5}\oplus\mathbf{\overline{5}}}
\nc{\vten}{\mathbf{10}\oplus\mathbf{\overline{10}}}
\nc{\zhol}{Z^{\rm hol}}
\nc{\xfb}{\,{\rm fb}}

\begin{document}

%
%

\preprint{UMN--TH--4110/22}
\preprint{FTPI--MINN--22/02}
\preprint{CERN-TH-2021-222}

\vspace*{1mm}

\title{Gravitational portals in the early Universe}

\author{Simon Cléry$^{a}$}
\email{simon.clery@ijclab.in2p3.fr}
\author{Yann Mambrini$^{a,b}$}
\email{yann.mambrini@ijclab.in2p3.fr}
\author{Keith A. Olive$^{c}$}
\email{olive@physics.umn.edu}
\author{Sarunas Verner$^{c}$}
\email{nedzi002@umn.edu}

\vspace{0.1cm}

\affiliation{
${}^a$ Universit\'e Paris-Saclay, CNRS/IN2P3, IJCLab, 91405 Orsay, France
 }
 \affiliation{
${}^b$ CERN, Theoretical Physics Department, Geneva, Switzerland
 }
 \affiliation{
${}^c$ 
 William I.~Fine Theoretical Physics Institute, 
       School of Physics and Astronomy,
            University of Minnesota, Minneapolis, MN 55455, USA
}

\begin{abstract} 

We consider the production of matter and radiation during reheating after inflation, restricting our attention solely to  gravitational interactions. 
Processes considered are the exchange 
of a graviton, $h_{\mu \nu}$, 
involved in the scattering of the inflaton or particles in the newly created radiation bath. In particular, we consider the gravitational production of dark matter (scalar or fermionic) from the thermal bath
as well as from scattering of the inflaton condensate. 
We also consider the gravitational production of radiation
from inflaton scattering. In the latter case, 
we also derive a {\it lower bound} on the maximal temperature 
of order of $10^{12}$ GeV for a typical $\alpha-$attractor scenario
from $\phi \phi \rightarrow h_{\mu \nu} \rightarrow$ Standard Model fields (dominated by the production of Higgs bosons). This
lower gravitational bound becomes the effective maximal
temperature for reheating temperatures, $\trh \lesssim 10^9$ GeV.
The processes we consider are all minimal in the sense that they
are present in {\em any} non-minimal extension of the Standard Model theory based on Einstein gravity and can not be neglected. 
We compare each of these processes to determine their relative importance in the production of both radiation and dark matter.

\end{abstract}

\maketitle




\section{Introduction}

Despite more than 80 years since the first indication of dark matter \cite{Zwicky:1933gu}, its nature and identity still remains a mystery \cite{book}. The hypothesis for a weakly interacting massive particle (WIMP) as a dark matter candidate is being challenged by an obvious lack of signal in dedicated direct detection experiments such as XENON1T \cite{XENON}, LUX \cite{LUX} or PANDAX \cite{PANDAX} (see \cite{Arcadi:2017kky} for a detailed review). These experiments exclude de facto a large part of the
parameter space in models where dark matter communicates with the Standard Model via the Higgs \cite{Burgess:2000yq,Higgsportal}, the $Z$ \cite{Zportal} or even an electroweak extension introducing a massive $Z'$ mediator \cite{Zpportal}. However, an alternative exists in the form of particles interacting very weakly with the thermal bath, and never having reached thermal equilibrium \cite{fimp}. The seclusion  can be justified by the weakness of a coupling (gravitational in the case of the gravitino \cite{nos,ehnos,kl,ekn,oss}) or by the exchange of very heavy mediators (generated by an extra $U(1)$ \cite{Bhattacharyya:2018evo}, moduli field \cite{fimpmoduli} or massive spin-2 field \cite{Bernal:2018qlk} as examples). A complete review can be found in \cite{Bernal:2017kxu} as well as related studies in \cite{Bernal,Bernal:2019mhf,Barman:2020plp,Chen:2017kvz}.

The minimal coupling one can imagine between dark matter and the Standard Model is gravitational mediated through a graviton \cite{ema,Redi:2020ffc}. As this coupling is unavoidable, any process invoking graviton exchange provides a lower limit on the amount of dark matter produced either via the thermal bath \cite{Garny:2015sjg,Tang:2017hvq,Chianese:2020yjo,Bernal:2018qlk,Redi:2020ffc,Bernal:2021kaj} or directly through the scattering of the inflaton \cite{MO,Barman:2021ugy}. The energy available in both cases partly compensates for the strong reduction in coupling by the Planck mass\footnote{Throughout the paper, we will consider the reduced Planck mass $M_P=1/\sqrt{8\pi G_N}\simeq 2.4 \times 10^{18}$ GeV.}, $M_P$. This is not too surprising. Indeed, we know that in the case of a FIMP, a coupling of the order of $\sim 10^{-11}$ is needed to produce dark matter in sufficient quantities. This corresponds to an effective coupling of the order of $\frac{E^2}{M_P^2}$, with $E \sim 10^{13}$ GeV representing the available energy  in the interaction. This energy corresponds, roughly, to the mass of the inflaton. It is therefore at the end of inflation, during the transition period between an inflaton-dominated universe and the radiative universe, called reheating, that the available energy is sufficient for the efficient gravitational production of dark matter.  

The reheating process is not instantaneous \cite{Giudice:2000ex,Bernal:2020gzm,GMOP}. The radiation bath may be produced by inflaton decays or
scattering which require a coupling of the inflaton to the Standard Model, or as we show below through the gravitational production of radiation. As the radiation begins to appear, 
the Universe rapidly achieves a maximum temperature, $\tmax$ and the reheating process continues until radiation domination is achieved at $\trh$. The evolution of the radiation density depends on \cite{GKMO1,GKMO2} 1) how it is produced, that is, through decays, or scatterings, 
2) the dominant final state particle spin, and 3) the form of the inflaton potential about its minimum, which we take as
 $V(\phi) \simeq \lambda \phi^k M_P^{4-k}$.
 This approximation is appropriate for the Starobinsky model \cite{Staro} (leading to $k=2$), as well as more general 
  $\alpha$-attractor type models \cite{Kallosh:2013hoa,Garcia:2021gsy}. Once the reheating is achieved, $T> \trh$, the inflaton disappears from the energy budget and the temperature evolves isentropically $T\propto a^{-1}$, where $a$ is the scale factor of the Universe. As we show below, the evolution of the radiation density can be modified by the gravitational production of Standard Model quanta which induces a lower bound on the maximum temperature of the Universe. We show that it is of the order of $10^{12}$ GeV, and is one of the main results of our work.

If the production of dark matter occurs during reheating, it is intimately linked to the behaviour of the inflaton and the evolution of the thermal bath.  Often it is assumed that the either the dark matter is directly coupled to the inflaton, in which case, it can be produced directly from inflaton decays \cite{egnop,grav2,GMOP,GKMO1} or it is coupled to the Standard Model, and thus produced thermal as the gravitino or other super-weakly interacting particles. In the latter case,
it has also been shown that radiative decay of the inflaton \cite{KMO} could be the dominant process to populate the dark Universe.

While reheating requires some coupling of the inflaton to the Standard Model (as will see gravitation interactions alone will not lead to radiation domination), the mechanism for producing dark matter may in fact be dominated solely by gravity.
In this paper, we analyze all processes involving a gravitational interactions, comparing the modes 
of production via the thermal bath, the scattering of the inflaton, and gravitational production of particles from the thermal bath
which subsequently produce dark matter through gravity as well. In this sense, each of the physical quantities we consider, such as the relic density or maximum temperature, must be considered as {\it lower bounds} as the gravitational process we compute are inevitable in any theory based on Einstein gravity. As a result, these lower bounds
must be taken into account in any kind of extension of the Standard Model, and can be thought of as a gravitational ``background noise". 
We do not consider preheating
via parametric or stochastic resonances as we did in \cite{GKMOV}, because we want to compute the minimal {\it unavoidable} amount of dark matter, 
and thus derive the strongest model-independent constraints on the dark matter mass, supposing that it only couples gravitationally.  

The only non-gravitational coupling we consider, is a coupling of the inflaton to SM fields to achieve reheating. Thus, we consider a generic Yukawa-like coupling of the form, $y \phi {\bar f} f$, where $f$ is some Standard Model fermion. We assume rapid thermalization, and these decays are (partially) responsible for the growing thermal bath. However the production of dark matter from the thermal bath is entirely gravitational. 

The paper is organized as follows.
The framework for our computation is outlined 
in Section~\ref{Sec:framework}. We consider
both scalar and fermionic dark matter
coupled to the Standard Model and the inflaton only through gravity. We compute the rates for the production of dark matter either through
thermal scattering (mediated by gravity alone)
or from the inflaton condensate. We choose an attractor form for the inflaton potential
which when expanded about its minimum, take the form $\phi^k$. Our results are sensitive to $k$. Reheating takes place as the inflaton oscillates about this minimum. In Section~\ref{Sec:dm} we consider three distinct gravitational process. The gravitation production of dark matter from the thermal bath; the gravitational production of dark matter from the condensate; and the gravitational production of the thermal bath from the condensate. We then compare each modes in Section~\ref{Sec:results}, 
before concluding in Section~\ref{Sec:conclusion}.

\section{The framework}
\label{Sec:framework}
\begin{figure}[ht]
\centering
\includegraphics[width=3.in]{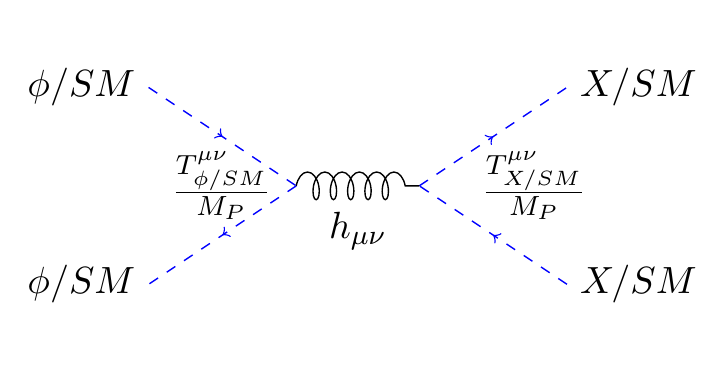}
\caption{\em \small Feynman diagram for the production of dark matter through the gravitational scattering of the Standard Model particle bath or inflaton condensate.
}
\label{Fig:feynman}
\end{figure}

We study universal gravitational interactions that must exist between the inflationary and dark sectors. 
If the space-time metric is expanded around flat space using $g_{\mu \nu} \simeq \eta_{\mu \nu} + {\tilde h}_{\mu \nu}$
the gravitational Lagrangian in the transverse-traceless gauge at second order can be written as
\beq
{\cal L} = \frac{M_P^2}{2} R \ni \frac{M_P^2}{8} (\partial^\alpha {\tilde h}^{\mu \nu})(\partial_\alpha {\tilde h}_{\mu \nu}) = \frac12 (\partial^\alpha h^{\mu \nu})(\partial_\alpha h_{\mu \nu})
\eeq
where $h_{\mu \nu} = (M_P/2) {\tilde h}_{\mu\nu}$ is the canonically normalized 
perturbation and and $M_P  = (8\pi G)^{-1/2} \simeq 2.4 \times 10^{18} \, \rm{GeV}$ is the reduced Planck mass.
Gravitational interactions are described by the Lagrangian~(see e.g., \cite{hol})
\beq
\sqrt{-g}{\cal L}_{\rm int}= -\frac{1}{M_P}h_{\mu \nu}
\left(T^{\mu \nu}_{SM}+T^{\mu \nu}_\phi + T^{\mu \nu}_{X} \right) \, .
\label{Eq:lagrangian}
\eeq
Here SM represents Standard Model fields, $\phi$ is the inflaton and $X$ is a dark matter candidate. The form of the stress-energy tensor $T^{\mu \nu}_i$ depends on the spin of the field, $i = 0, \, 1/2, \, 1$,\footnote{In this work we consider dark matter candidates which are either real scalars or a Dirac fermion.} and is given by
\bea
T^{\mu \nu}_{0} &=&
\partial^\mu S \partial^\nu S-
g^{\mu \nu}
\left[
\frac{1}{2}\partial^\alpha S \partial_\alpha S-V(S)\right] \, ,
\label{Eq:tensors}
\\
T^{\mu \nu}_{1/2} &=&
\frac{i}{4}
\left[
\bar \chi \gamma^\mu \overset{\leftrightarrow}{\partial^\nu} \chi
+\bar \chi \gamma^\nu \overset{\leftrightarrow}{\partial^\mu} \chi \right] \nonumber \\
&& -g^{\mu \nu}\left[\frac{i}{2}
\bar \chi \gamma^\alpha \overset{\leftrightarrow}{\partial_\alpha} \chi
-m_\chi \bar \chi \chi\right] \, , \\
\label{Eq:tensorf}
T_{1}^{\mu \nu} &=& \frac{1}{2} \left[ F^\mu_\alpha F^{\nu \alpha} + F^\nu_\alpha F^{\mu \alpha} - \frac{1}{2} g^{\mu \nu} F^{\alpha \beta} F_{\alpha \beta} \right] \, ,
\label{Eq:tensorv}
\eea
where $V(S)$ is the scalar potential for either the 
scalar dark matter or inflaton, with $S \; = \; X, \phi$, and 
$F_{\mu \nu}=\partial_\mu A_\nu-\partial_\nu A_\mu$
is the field strength for a vector field, $A_\mu$. In Fig.~\ref{Fig:feynman}, we show the $s$-channel exchange of a graviton obtained from the Lagrangian~(\ref{Eq:lagrangian}) for the production of dark matter from either the inflaton condensate or Standard Model fields. In addition, a similar diagram exists
for the production of Standard Model fields (during the reheat process) from the inflaton condensate in the initial state. 

Although the direct coupling to the massless graviton appears to be feeble due to Planck suppression, the energy available in the thermal bath during the initial stage of reheating is large enough to make the gravitational production rates significant.

The scattering amplitudes related to the production rate of the processes $\phi/{\rm{SM}}^i(p_1)+\phi/{\rm{SM}}^i(p_2) \rightarrow {\rm{SM}}^i/X^j (p_3)+{\rm{SM}}^i/X^j(p_4)$ can be parametrized by
\begin{equation}
\mathcal{M}^{i j} \propto M_{\mu \nu}^j \Pi^{\mu \nu \rho \sigma} M_{\rho \sigma}^i \;, 
\end{equation}
where ($i$, $j$) denotes the spin of the (initial,final) state  involved in the scattering process and $i,j=0,1/2,1$. $\Pi^{\mu \nu \rho \sigma}$ is the graviton propagator for the canonical field $h$ with momentum $k = p_1+p_2$,
\begin{equation}
 \Pi^{\m\n\rho\sigma}(k) = \frac{\eta^{\rho\n}\eta^{\sigma\m} + 
\eta^{\rho\m}\eta^{\sigma\n} - \eta^{\rho\sigma}\eta^{\m\n} }{2k^2} \, .
\end{equation} 
    The partial amplitudes, $M_{\mu \nu}^i$, are given by
\bea 
M_{\mu \nu}^{0} &=& \frac{1}{2}\left[p_{1\mu} p_{2\nu} + p_{1\nu} p_{2\mu} - \eta_{\mu \nu}p_1\cdot p_2 - \eta_{\mu \nu} V''(S)\right] \,, \\ 
M_{\mu \nu}^{1/2} &=&  \frac{1}{4} {\bar v}(p_2) \left[ \gamma_\mu (p_1-p_2)_\nu + \gamma_\nu (p_1-p_2)_\mu \right] u(p_1) \, , \\
M_{\mu \nu}^{1} &=& \frac{1}{2} \bigg[ \epsilon_{2}^{*} \cdot \epsilon_{1}\left(p_{1 \mu} p_{2 \nu}+p_{1 \nu} p_{2 \mu}\right)
\nonumber\\
&-&\epsilon_{2}^{*} \cdot p_{1}\left(p_{2 \mu} \epsilon_{1 \nu}+\epsilon_{1 \mu} p_{2 \nu}\right) - \epsilon_{1} \cdot p_{2}\left(p_{1 \nu} \epsilon_{2 \mu}^{*}+p_{1 \mu} \epsilon_{2 \nu}^{*}\right)
\nonumber\\
&+&p_{1} \cdot p_{2}\left(\epsilon_{1 \mu} \epsilon_{2 \nu}^{*}+\epsilon_{1 \nu} \epsilon_{2 \mu}^{*}\right)  \nonumber \\
&+&g_{\mu \nu}\left(\epsilon_{2}^{*} \cdot p_{1} \epsilon_{1} \cdot p_{2}-p_{1} \cdot p_{2} \, \epsilon_{2}^{*} \cdot \epsilon_{1}\right) \bigg]  \, ,
\label{partamp}
\eea
with analogous expressions for dark matter in terms of the dark matter momenta, $p_3, p_4$, and potential $V(X)$, if $X$ is a scalar. For an initial state inflaton with $S = \phi$, we replace $M^0_{\mu\nu}$ with $T^0_{\mu \nu}$ from Eq.~(\ref{Eq:tensors}). As we only consider vectors in the Standard Model, their masses have been neglected in Eq.~(\ref{partamp}).

In what follows, we consider three distinct processes based on the diagram in Fig.~\ref{Fig:feynman}: for the production of dark matter, A) SM + SM $\to X + X$; B) $\phi + \phi \to X + X$, where the latter involves the inflaton condensate (zero mode) in the initial state rather than an initial state particle with momentum $p_{1,2}$ (see below for more detail), and C) $\phi + \phi \to$ SM + SM, as a minimal and unavoidable contribution to the reheating process. 

The dark matter production rate from SM fields can be readily calculated by assuming that the initial particle states are massless. This assumption can be justified by the fact that the energy associated with the momenta, $p_1 \, , p_2$ is extremely large
at the end of inflation and dominates over electroweak scale quantities.

The dark matter production rate $R(T)$ for the 
SM+SM $\rightarrow$ $X + X$ process with amplitude $\cal{M}$~\footnote{It should be noted that we include the symmetry factors associated with identical initial and final states in the squared amplitude, $|{\cal{M}}|^2$.} is
\beq
R(T) = \frac{1}{1024 \pi^6}\int f_1 f_2 E_1 \diff E_1 E_2 \diff E_2 \diff \cos \theta_{12}\int |{\cal M}|^2 \diff \Omega_{13} \, ,
\eeq
where $E_i$ denotes the energy of particle $i=1,2,3,4$,
$\theta_{13}$ and $\theta_{12}$ are the angles formed by momenta ${\bf{p}}_{1,3}$ and ${\bf{p}}_{1,2}$, respectively, and 
\beq
f_{i}= \frac{1}{e^{E_i/T}\pm 1} \, ,
\eeq
represent the assumed thermal distributions of the incoming SM particles.

The total amplitude squared for the gravitational scattering process SM+SM $\rightarrow$ $X_j + X_j$ is given by a sum of the three amplitudes associated with different initial state spins,
\begin{equation}
    \label{Eq:ampscat}
    |\mathcal{M}|^{2}=4 |\mathcal{M}^{0}|^{2}+45|\mathcal{M}^{1 / 2}|^{2}+12|\mathcal{M}^{1}|^{2} \, .
\end{equation}
These were calculated in \cite{Bernal:2018qlk} and it was found that the dark matter production rate is given by
\beq
R^T_j=R_j(T)= \beta_j\frac{T^8}{M_P^4} \, ,
\label{Eq:ratethermal}
\eeq
where $j$ refers to the spin of $X$ (either 0 or 1/2), the constants $\beta_j$ and details related to the computation of dark matter production rate and the amplitude squared are given in Appendix~\ref{sec:appendixA}.

For the production of dark matter through the scattering of the inflaton condensate we consider the time-dependent oscillation of a classical inflaton field $\phi(t)$. 
Since our computation depends explicitly on inflaton potential, we consider the $\alpha$-attractor T-model~\cite{Kallosh:2013hoa} as a specific example, 
\begin{equation}
    V(\phi) \; = \;\lambda M_P^{4}\left|\sqrt{6} \tanh \left(\frac{\phi}{\sqrt{6} M_P}\right)\right|^{k} \, ,
\label{Vatt}
\end{equation}
which can be expanded about the origin\footnote{It should be noted that our discussion is general and not limited to T-models of inflation.}
\begin{equation}
    \label{Eq:potmin}
    V(\phi)= \lambda \frac{\phi^{k}}{M_{P}^{k-4}}, \quad \phi \ll M_{P} \, .
\end{equation}
The time-dependent oscillating inflaton field can be parametrized as
\begin{equation}
    \label{Eq:oscillation}
    \phi(t) \; = \; \phi_0(t) \cdot \mathcal{P}(t) \, , 
\end{equation}
where $\phi_0(t)$ is the time-dependent amplitude that includes the effects of redshift and $\mathcal{P}(t)$ describes the periodicity of the oscillation.

To calculate the dark matter production rate, we combine the potential~(\ref{Eq:potmin}) with Eq.~(\ref{Eq:oscillation}), which leads to $V(\phi) = V(\phi_0) \cdot \mathcal{P}(t)^k$. We next expand the potential energy in terms of the Fourier modes \cite{Ichikawa:2008ne,Kainulainen:2016vzv,GKMO2}
\beq
V(\phi)=V(\phi_0)\sum_{n=-\infty}^{\infty} {\cal P}_n^ke^{-in \omega t}
=\rho_\phi\sum_{n = -\infty}^{\infty} {\cal P}_n^ke^{-in \omega t} \, ,
\eeq
where $\omega$ is the frequency of oscillation of $\phi$, given by \cite{GKMO2}
\beq
\label{eq:angfrequency}
\omega=m_\phi \sqrt{\frac{\pi k}{2(k-1)}}
\frac{\Gamma(\frac{1}{2}+\frac{1}{k})}{\Gamma(\frac{1}{k})}.
\eeq

For scalar dark matter, we find that the particle production rate per unit volume and unit time for an arbitrary value of $k$ is given by
\beq
\label{Eq:ratephi0}
R^{\phi^k}_0=\frac{2 \times \rho_\phi^2}{16 \pi M_P^4} \Sigma_0^k \, ,
\eeq
where the factor of two accounts for the fact we produce two dark matter particles per scattering, with
\begin{equation}
   \Sigma_0^k = \sum_{n = 1}^{\infty}  |{\cal P}^k_n|^2\left[1+\frac{2m^2_X}{E_n^2}\right]^2 
\sqrt{1-\frac{4m_X^2}{E_n^2}} \, ,
\label{Sigma0k}
\end{equation}
where $E_n = n \omega$ is the energy of the $n$-th inflaton oscillation mode and $m_X$ is the mass of the produced dark matter. A detailed calculation of this rate is presented in Appendix~\ref{sec:appendixB}.

For the case $k = 2$, we find that the particle production rate is given by
\beq
\label{eq:ratescalar2}
R^{\phi^2}_0=\frac{ 2 \times \rho_\phi^2}{256 \pi M_P^4}
\left(1+\frac{m_X^2}{2m^2_\phi}\right)^2\sqrt{1-\frac{m_X^2}{m_\phi^2}} \, ,
\eeq
where $m_{\phi}^2 = V''(\phi_0)$, and since $\sum \mathcal{P}^2_n = \cos^2(m_{\phi} t)$, we find that only the second Fourier mode in the sum contributes, with $\sum |\mathcal{P}^2_n|^2 = \frac{1}{16}$ and $E_2 = 2m_{\phi}$.\footnote{We note that the rate calculated here differs from~\cite{MO} by a factor of 8,  because in the latter the inflaton was treated as a particle and not a condensate resulting in a difference by a factor of 2 in the applied symmetry factors. In addition, the interaction considered there did not use a properly normalized graviton resulting in a factor of 2 in the vertex and 16 in the rate.}

For a fermionic dark matter candidate, we find the following rate
\beq
\label{eq:rateferm}
R^{\phi^k}_{1/2}=\frac{2 \times \rho_\phi^2}{4 \pi M_P^4}
\frac{m_X^2}{m_\phi^2}
\Sigma_{1/2}^k \, ,
\eeq
where the factor of two accounts for the  sum over the particle and antiparticle final states, with
\begin{equation}
    \label{eq:ratefermion2}
    \Sigma_{1/2}^k = \sum_{n=1}^{+\infty} |{\cal P}_n^k|^2
    \frac{m_\phi^2}{E_n^2}
\left[1-\frac{4 m_X^2}{E_n^2}\right]^{3/2} \, .
\end{equation}
For the case $k = 2$, we obtain
\beq
R^{\phi^2}_{1/2}\simeq\frac{2 \times \rho_\phi^2}{256 \pi M_P^4}\frac{m_X^2}{m^2_\phi}
\left[1-\frac{m_X^2}{m_\phi^2}\right]^{3/2} \, .
\eeq
A detailed discussion related to the dark matter production rates through the inflaton condensate scattering is given in Appendix~\ref{sec:appendixB}.
    
For the production of SM fields from inflaton oscillations, 
we follow the same procedure, but replace the partial amplitude, $M_{\mu \nu}^j$, for dark matter with the appropriate amplitude
involving SM fields. Below, we consider only the example of producing Higgs bosons, namely $\phi + \phi \to H + H$.

\section{Gravitational production of quanta}
\label{Sec:dm}

As we discussed in the previous section, the graviton can act as a portal between the inflaton, SM fields and a potential dark matter candidate. As outlined above we here consider three cases in detail:

\begin{enumerate}[A.]
\item{The graviton portal between a thermal bath and dark matter. This is essentially a gravitational freeze-in mechanism for the production of dark matter. }
\item{The graviton portal between the inflaton and dark matter. In this case, the inflaton  directly populates the dark matter without the need of either the thermal bath or a mediator between the SM and the dark matter candidate.}
\item{The graviton portal between the inflaton and the Standard Model sector to produce a radiative bath at the start of reheating.}
\end{enumerate}

\subsection{${\rm SM~SM} \rightarrow h_{\mu \nu} \rightarrow {\rm DM~DM}$}

The spin-2 portal for the production of dark matter was considered recently in \cite{Bernal:2018qlk} for both massive and massless spin-2 fields. Here we restrict our attention to 
the massless (graviton) portal. For an inflaton potential with $k=2$, the scattering cross section between SM fields and dark matter is proportional to $T^2/M_P^4$, and we expect the resulting dark matter abundance to be primarily sensitive to the reheating temperature (rather than the maximum temperature attained during the reheating process). Sensitivity to $\tmax$ requires a cross section with a steep dependence on temperature, $\sigma \propto T^n$, with $n \ge 6$. When $k>2$, sensitivity to $\tmax$ requires only $n > (10-2k)/(k-1)$ when the primary reheating mechanism is determined by inflaton decays as discussed below. Then, for example, when $k=4$, when
$n>2/3$,  the dark matter abundance becomes sensitive to $\tmax$. For the graviton portal, then, this occurs when $k\ge 3$.

The gravitational scattering of particles in the primordial plasma can produce
massive particles playing the role of a viable dark matter candidate $X$.
Then, the matter density $n_X$ obeys the classical Boltzmann equation\footnote{We note that we include the relevant factors of 2 associated with identical initial states in the definition of the particle production rate.}
\beq
\frac{dn_X}{dt} + 3Hn_X = R^T_X \,,
\label{Eq:boltzmann1}
\eeq
where $H=\frac{\dot a}{a}$ is the Hubble parameter. 
It is more convenient to work with $a$ as dynamical parameter,
rather than $t$ or $T$. 
Eq.(\ref{Eq:boltzmann1}) can then be rewritten
\beq
\frac{dn_X}{da}+ 3\frac{n_X}{a} = \frac{R^T_X(a)}{H a} \, .
\eeq
Since the production rate $R^T_X$ is dependent on the initial state energies, $i.e.$
 of the temperature of the thermal bath, one needs 
 the expression of $T(a)$ to solve
the Boltzmann equation in terms of the scale factor. 
We explain the functional dependence of $R^T_X$ on $a$ below.
Defining the comoving number $Y_X= n  a^3$, we obtain
\beq
\frac{dY_X}{da}=\frac{a^2R^i_X(a)}{H} \, .
\label{Eq:boltzmann2}
\eeq

We assume an inflaton potential of the form given in Eq.~(\ref{Eq:potmin}).   We next apply the expressions for
 energy conservation for the inflaton density $\rho_\phi$ and 
the radiation density $\rho_R$
\bea
&&
\frac{d \rho_\phi}{dt} + 3H(1+w_{\phi})\rho_\phi \simeq -(1+w_{\phi})\Gamma_\phi \rho_\phi 
\label{Eq:diffrhophi}
\\
&&
\frac{d \rho_R}{dt} + 4H\rho_R \simeq (1+w_{\phi})\Gamma_\phi \rho_\phi \, .
\label{Eq:diffrhor}
\eea
where $w_{\phi} = \frac{P_{\phi}}{\rho_{\phi}}=\frac{k-2}{k+2}$ \cite{GKMO2} is the equation of state parameter.
Here we assume that reheating primarily occurs due to the inflaton effective coupling to the Standard Model fermions, given by the Lagrangian
\begin{align}
\mathcal{L}^y_{\phi-SM}  =  - y \phi {\bar f} f \, ,
 \label{phidecaycoupling}
\end{align}
where $y$ is a Yukawa-like coupling and $f$ is a Standard Model fermion. The width of $\phi$ is easily determined from the coupling (\ref{phidecaycoupling})
\beq
\Gamma_\phi=\frac{y^2}{8 \pi}m_\phi \, .
\eeq
Note that for $k > 2$, $m_\phi$ depends on $\phi$ and hence on the scale factor $a$. We
defined the inflaton energy density and pressure as
\beq
\rho_\phi = \frac{1}{2}\langle \dot \phi^2 \rangle
+\langle V(\phi) \rangle,
~~~~P_\phi=\frac{1}{2}\langle \dot \phi^2 \rangle 
- \langle V(\phi) \rangle.
\eeq
We can solve Eqs.~(\ref{Eq:diffrhophi}, \ref{Eq:diffrhor})
and obtain \cite{GKMO1,GKMO2}
\beq
\rho_\phi(a) = \rho_{\rm end} \left(\frac{a_{\rm end}}{a} \right)^\frac{6k}{k+2}
\label{rhophia}
\eeq
and
\beq
\rho_R(a)=\rhorh\left(\frac{\arh}{a}\right)^{\frac{6k-6}{k+2}}
\frac{1-\left(\frac{\ae}{a}\right)^{\frac{14-2k}{k+2}}}
{1-\left(\frac{\ae}{\arh}\right)^{\frac{14-2k}{k+2}}} \, ,
\label{Eq:rhoR}
\eeq
where these relations hold for $a_{\rm end} \ll a \ll a_{\rm{RH}}$. $a_{\rm end}$ is a reference point marking the end of inflation.  $\rho_\phi(a_{\rm end})$ corresponds to the total energy density (there is virtually no radiation at this point)
when the slow-roll parameter $\epsilon = 1$. At this moment, $\rho_{\rm end} = \frac32 V(\phi_{\rm end})$ \cite{egno5}.
Note that this solution possesses a maximum for $\rho_R(a)$ (at $a = a_{\rm max}$).
We have also defined $\rhorh$ and $a_{\rm RH}$
such that $\rho_R(a_{\rm RH})=\rho_\phi(a_{\rm RH})$. 
Since
\beq
\rho_R=\frac{g_T\pi^2}{30}T^4\equiv \alpha T^4 \, ,
\eeq
where $g_T$ is
the number of relativistic degrees of freedom at the temperature, $T$. Thus, we have $\rho_R(a_{\rm max}) = \alpha \tmax^4$ and $\rho_R(a_{\rm RH}) = \alpha \trh^4$.  
The ratio of $a_{\rm max}$ to $a_{\rm end}$ is fixed and depends only on $k$ \cite{GKMO1}
\beq
\frac{a_{\rm max}}{\ae}=\left(\frac{2k+4}{3k-3}\right)^{\frac{k+2}{14-2k}} \, .
\label{Eq:amax}
\eeq

Since we can express $T$ as function of the scale factor, $a$, with Eq.~(\ref{Eq:rhoR}), we can implement that relation in Eq.~(\ref{Eq:ratethermal}) to obtain $R_X^T$ as function of $a$,
\beq
R_X^T(a)=\beta_X \frac{\rhorh^2}{\alpha^2M_P^4} \left(\frac{\arh}{a}\right)^{\frac{12k-12}{k+2}} 
\left[\frac{1 - \left(\frac{a_{\rm end}}{a} \right)^\frac{14-2k}{k+2}}
{1 - \left(\frac{a_{\rm end}}{\arh} \right)^\frac{14-2k}{k+2}}
\right]^2. 
\label{rateXT}
\eeq
Using $H \simeq \frac{\sqrt{\rho_\phi(a)}}{\sqrt{3}M_P}$,
which is valid for $a \ll a_{\rm RH}$,  Eq.~(\ref{Eq:boltzmann2})
becomes
\beq
\frac{dY_X}{da}=\frac{\sqrt{3}M_P}{\sqrt{\rhorh}}a^2\left(\frac{a}{a_{\rm RH}}\right)^{\frac{3k}{k+2}}R_X^T(a) \, .
\label{Eq:boltzmann3}
\eeq
The solution to this equation is 
\begin{widetext}
\begin{eqnarray}
n^T_X(a_{\rm RH}) & = &
\frac{\beta_X\sqrt{3}}{\alpha^2M_P^3}
\frac{\rhorh^{3/2}}{(1-(\ae/\arh)^{\frac{14-2k}{k+2}})^2} 
\begin{cases}
\frac{k+2}{6} \left(\frac{1}{3-k} + \frac{3}{k-1}\left(\frac{a_{\rm end}}{a_{\rm RH}} \right)^\frac{14-2k}{k+2}  +  \frac{(k-7)^2}{k^3+k^2-17k+15}\left(\frac{a_{\rm end}}{a_{\rm RH}} \right)^\frac{18-6k}{k+2} - \frac{3}{k+5}\left(\frac{a_{\rm end}}{a_{\rm RH}} \right)^\frac{28-4k}{k+2}\right)  \\
\qquad \qquad \qquad \qquad \qquad \qquad \qquad \qquad \qquad \qquad \qquad   [k\ne 3]  \\
\\
\\
  \ln\left(\frac{a_{\rm RH}}{a_{\rm end}} \right) - 
  \frac{5}{16} \left( 3 -4\left(\frac{a_{\rm end}}{a_{\rm RH}} \right)^\frac85 + \left(\frac{a_{\rm end}}{a_{\rm RH}} \right)^\frac{16}{5} \right)\qquad \qquad [k=3]
\end{cases}
\label{Eq:nxthermal}
\end{eqnarray}
\end{widetext}
where we integrated Eq.~(\ref{Eq:boltzmann3}) between the values of the scale factor corresponding to the end of inflation, $a_{\rm end}$,  and the reheating temperature 
(reached at $a_{\rm RH}$).

Writing the 
relic abundance \cite{book}
\beq
\Omega_Xh^2 = 1.6\times 10^8\frac{g_0}{g_{\rm RH}}\frac{n(\trh)}{\trh^3}\frac{m_X}{1~{\rm GeV}},
\eeq
and inserting Eq.~(\ref{Eq:nxthermal}), we obtain
\begin{widetext}
\bea
    \Omega^T_X h^2 = \Omega_k \times
    \begin{cases}
 \frac{k+2}{6} \left(\frac{1}{3-k}+ \frac{3}{k-1}\left(\frac{\rho_{\rm RH}}{\rho_{\rm end}} \right)^\frac{7-k}{3k} + \frac{(k-7)^2}{k^3+k^2-17k+15}\left(\frac{\rho_{\rm RH}}{\rho_{\rm end}} \right)^\frac{3-k}{k} - \frac{3}{k+5}\left(\frac{\rho_{\rm RH}}{\rho_{\rm end}} \right)^\frac{14-2k}{3k}\right)~~~~~~~[k\ne3]
\\
\\
\frac{5}{18}\ln\left(\frac{\rho_{\rm end}}{\rho_{\rm RH}} \right) -\frac{5}{16} \left( 3 -4\left(\frac{\rho_{\rm RH}}{\rho_{\rm end}} \right)^\frac49 + \left(\frac{\rho_{\rm RH}}{\rho_{\rm end}} \right)^\frac{8}{9} \right)~~~~~~~~~~~~~~~~~~~~~~~~~~~~~~~~~~~~~~~~~~~~~~[k=3]

\end{cases} \, 
    \label{22}
\eea
\end{widetext}
with 
\beq
\Omega_k = 1.6 \times 10^8 \frac{g_0}{g_{\rm RH}}
    \frac{\beta_X\sqrt{3}}{\sqrt{\alpha} }
    \frac{m_X}{1~\rm{GeV}}
    \frac{\trh^3}{M_P^3}\left[1-\left(\frac{\rhorh}{\rhoe}\right)^{\frac{7-k}{3k}}\right]^{-2} \, ,
\eeq
where $g_0 = 43/11$ and we take $g_{\rm RH} = 427/4$ as the Standard Model value.

We observe that, for a given reheating temperature, the relic abundance decreases 
with $k$. Furthermore, whereas $\Omega_X^T h^2 \propto \trh^3$ for a quadratic potential, it
becomes $\propto \trh^2$ for a quartic potential, and even $\propto \trh$ for 
$k=6$. This comes from the fact that the Hubble parameter, dominated by the evolution of the inflaton, has a greater dependence on $T$ for larger values of $k$, slowing 
down the production mechanism for large temperatures.

\subsection{$\phi~\phi \rightarrow h_{\mu \nu} \rightarrow {\rm DM~DM}$}

As noted earlier, it is also possible that the inflaton condensate can lead to the direct production of dark matter through single graviton exchange \cite{MO}. Here, we generalize that result for $k \ge 2$. Having computed the production rate in Eqs.~(\ref{Eq:ratephi0}) and (\ref{eq:rateferm}) for 
scalar and fermionic dark matter respectively, we can replace $R^T_X$ with the rates in Eq.~(\ref{Eq:boltzmann3}).
Then integrating 
\beq
\frac{dY_X}{da}=\frac{\sqrt{3}M_P}{\sqrt{\rhorh}}a^2\left(\frac{a}{a_{\rm RH}}\right)^{\frac{3k}{k+2}}R_X^{\phi^k}(a)
\label{Eq:boltzmann4}
\eeq
between $a_{\rm end}$ and $a_{\rm RH}$ gives for scalar dark matter
\beq
n_0^\phi(a_{\rm RH})=\frac{\sqrt{3}\rhorh^{3/2}}{8 \pi M_P^3}
\frac{k+2}{6k-6}
\left[
\left(\frac{a_{\rm RH}}{a_{\rm end}}\right)^{\frac{6k-6}{k+2}}-1
\right]\Sigma^k_0
\label{Eq:nsphi}
\eeq
which can be expressed as function of $\rhoe$ 
using Eq.~(\ref{rhophia}):
\beq
n_0^\phi(a_{\rm RH})\simeq
\frac{\sqrt{3}\rhorh^{3/2}}{8 \pi M_P^3}
\frac{k+2}{6k-6}
\left(\frac{\rhoe}{\rhorh}\right)^{1-\frac{1}{k}}\Sigma^k_0,
\label{n0phi}
\eeq
or
\bea
\frac{\Omega_0^\phi h^2}{0.1}&\simeq& \left(\frac{\rhoe}{10^{64} {\rm GeV}^4}\right)^{1-\frac{1}{k}}
\left(\frac{10^{40}{\rm GeV}^4}{\rhorh}\right)^{\frac{1}{4}-\frac{1}{k}}
\left(\frac{k+2}{6k-6}\right)
\nonumber
\\
&
\times&
\Sigma_0^k\times \frac{m_X}{ 2.4\times 10^{\frac{24}{k}-7} {\rm GeV}}
\label{Eq:omega0}
\eea
where we assumed $a_{\rm RH} \gg a_{\rm end}$. Note that the dependence on $\rho_\phi$ used in Eq.~(\ref{Eq:boltzmann4})
hides the fact that we considered a decaying inflaton during the reheating.  

For fermionic dark matter we obtained
\bea
&&
n_{1/2}^\phi(a_{\rm RH})=
\frac{m_X^2\sqrt{3}(k+2)\rhorh^{\frac{1}{2}+\frac{2}{k}}}{12 \pi k(k-1)\lambda^{\frac{2}{k}}M_P^{1+\frac{8}{k}}}
\left[\left(\frac{a_{\rm RH}}{a_{\rm end}}\right)^{\frac{6}{k+2}}
-1\right]\Sigma_{\frac{1}{2}}^k
\nonumber
\\
&&
\simeq 
\frac{m_X^2\sqrt{3}(k+2)\rhorh^{\frac{1}{2}+\frac{2}{k}}}{12 \pi k(k-1)\lambda^{\frac{2}{k}}M_P^{1+\frac{8}{k}}}
\left(\frac{\rhoe}{\rhorh}\right)^{\frac{1}{k}}
\Sigma_{\frac{1}{2}}^k
\label{Eq:nhalf}
\eea
where we used
\beq
m_\phi^2=V''(\phi_0)=k(k-1)\lambda^{\frac{2}{k}}M_P^2
\left(\frac{\rho_\phi}{M_P^4}\right)^{1-\frac{2}{k}}.
\eeq
We can simplify the expression to write
\bea
\frac{\Omega_{1/2}^\phi h^2}{0.1}&=&
\frac{\Sigma_{1/2}^k}{2.4^{\frac{8}{k}}}\frac{k+2}{k(k-1)}
\left(\frac{10^{-11}}{\lambda}\right)^{\frac{2}{k}}
\left(\frac{10^{40} {\rm GeV}^4}{\rhorh}\right)^{\frac{1}{4}-\frac{1}{k}}
\nonumber
\\
&\times&
\left(\frac{\rhoe}{10^{64} {\rm GeV}^4}\right)^{\frac{1}{k}}
\left(\frac{m_X}{8.3\times 10^{6+\frac{6}{k}} {\rm GeV}}\right)^3
\label{Eq:omegaphihalf}
\eea

Up until now, we have assumed that the thermal bath was produced via inflaton decays. 
However, for low reheat temperatures, and hence
small values of the Yukawa-like inflaton coupling, $y$, it is possible that radiation,
in the form of Higgs bosons, is produced 
directly from the condensate via gravitational interactions.  This is considered in the next subsection.

\subsection{$\phi~ \phi \rightarrow h_{\mu \nu} \rightarrow {\rm SM~SM }$}

The calculation for the production of SM fields produced
by the scattering of the inflaton via gravity is similar to the preceding calculation for dark matter.
As was shown in \cite{MO} and \cite{GKMO2}, there exists the possibility that the thermal bath is produced not by inflaton decay but rather by inflaton scattering after inflation. This occurs for instance for low values of $y$. In this case, the maximum temperature is not given by the inflaton width, but by the scattering process, whereas the final reheating (and thus $\trh$) is still dominated by the decay.
This is illustrated in Fig.~\ref{Fig:rhoR_num} below. In fact, the gravitational scattering $\phi \phi \rightarrow h_{\mu \nu} \rightarrow HH$ is always present and can never be eliminated. Such a process  generates an effective coupling
\beq
{\cal L}_h = \sigma_h \phi^2 H^2 \, .
\eeq
From Eq.~(A.23) of \cite{GKMO2},
we can write the left-hand side of Eq.~(\ref{Eq:diffrhor}) as 
\beq
(1+w)\Gamma_\phi \rho_\phi = N \frac{\sigma_h^2}{4 \pi} \phi_0^4 \omega
\sum_{n=1}^\infty n |{\cal P}_n^k|^2.
\label{Eq:rsigmah}
\eeq
where $N=4$ is the number of real scalars in the Standard Model, when we neglect the Higgs mass. Identifying this rate with that in Eq.~(\ref{Eq:ratephi0}),
and $ (1+w)\Gamma_\phi \rho_\phi = \omega R_0^{\phi^k}$,  we deduce that
\beq
\sigma_h=\frac{\rho_\phi}{8 M_P^2 \phi_0^2},
\label{quartic}
\eeq
for each real scalar. 
Thus for the Standard Model Higgs, and in  the case $k=2$ we have
\beq
\sigma_h=\frac{m_\phi^2}{16 M_P^2} \simeq 9.8 \times 10^{-12} \left(\frac{m_\phi}{3 \times 10^{13}~{\rm GeV}} \right)^2.
\eeq
$\sigma_h$ can be considered as the lowest possible and inevitable value for a quartic coupling between the inflaton and scalars.
This may be important and even dominate  the reheating process at its earliest stages. We note that in a theory with additional weak scale scalars such as the minimal supersymmetric Standard Model (MSSM), the gravitational production is increased due to the large number of scalars, $N=98$ in the MSSM. 
Note also that there is a minimal gravitational production rate for the production of SM fermions and gauge bosons though this is completely negligible due to the mass suppression (see e.g. Eq.~(\ref{eq:rateferm}) for fermions). Thus if we restrict
our attention to the Standard Model, we take
$N =4$ corresponding to the four real scalar degrees of freedom.

We now recompute the evolution of the radiation density using
Eq.~(\ref{Eq:diffrhor}) and (\ref{Eq:rsigmah}),
\beq
\frac{d \rho^h_R}{dt} + 4 H \rho^h_R = N \frac{\rho_\phi^2 \omega}{16 \pi M_P^4}
\sum_{n=1}^\infty n |{\cal P}_n^k|^2 \, .
\label{Eq:rhorsigma}
\eeq
The solution of (\ref{Eq:rhorsigma}) is 
\begin{eqnarray}
\rho^h_R&=&N\frac{\sqrt{3}M_P^4\gamma_k\Sigma^h_k}{16 \pi}
\left(\frac{\rho_e}{M_P^4}\right)^{\frac{2k-1}{k}}
\frac{k+2}{8k-14} \nonumber \\
&& \times \left[\left(\frac{a_e}{a}\right)^4
-\left(\frac{a_e}{a}\right)^{\frac{12k-6}{k+2}}\right]
\label{Eq:rhorsigma_sol}
\end{eqnarray}
with
\beq
\gamma_k=\sqrt{\frac{\pi}{2}}k
\frac{\Gamma(\frac{1}{2}+\frac{1}{k})}{\Gamma(\frac{1}{k})}
\lambda^{\frac{1}{k}}
\eeq
and
\beq
\Sigma^h_k=\sum_{n=1}^\infty n |{\cal P}_n^k|^2 \, .
\label{Eq:sumhk}
\eeq

Once again, there is a maximum temperature which can be determined by from the value of $a_{\rm end}/a$ which maximizes Eq.~(\ref{Eq:rhorsigma_sol}),
\beq
\frac{\ae}{a_{\rm max}}=\left( \frac{2k + 4}{6k-3}\right)^{\frac{k+2}{8k-14}},
\eeq
and hence a maximum radiation density,
\beq
\rho^h_{\rm max}=N \frac{\sqrt{3}M_P^4\gamma_k\Sigma^h_k}{16 \pi}
\left(\frac{\rhoe}{M_P^4}\right)^{\frac{2k-1}{k}}\frac{k+2}{12k-6}
\left(\frac{2k+4}{6k-3}\right)^{\frac{2k+4}{4k-7}}
\label{Eq:rhomaxbis}
\eeq
\\
For $k=2$ we have
\beq
T_{\rm max}^h\simeq 3.0 \times 10^{12}
\left(\frac{\rhoe}{10^{64}~{\rm GeV}^4}\right)^{\frac{3}{8}} \left(\frac{m_\phi}{3\times 10^{13}~ {\rm GeV}} \right)^{\frac14}
{\rm GeV} \, ,
\label{Eq:tmax}
\eeq
where we have taken $N=4$ and scales as $N^{1/4}$. 
Furthermore, the sum $\Sigma^h_k$
(\ref{Eq:sumhk}) begins at $n=2$, because 2 modes scatter, and the 
initial mode has an energy of $2\omega$, which implies for $k=2$, 
\beq
\Sigma^h_2= 2\times |{\cal P}_2^2|^2=2\times \frac{1}{16}=\frac{1}{8}\, .
\eeq

It is important to stress the importance of Eqs.~(\ref{Eq:rhomaxbis}) and (\ref{Eq:tmax}). These correspond to an absolute  
{\it lower bound} on the maximal temperature of the Universe.
We have not made any assumption other than the existence of a complex Higgs doublet and the inflaton coupled only through gravity. Our calculation implies that the Universe must have passed through this (or a higher) temperature during the early stages of reheating.

For $k=2$, the radiation density produced by inflaton scattering as computed above never comes to dominate the energy density and can not lead to reheating. Although scattering can lead to reheating if $k \ge 4$ \cite{GKMO2}.
Gravitational scattering is less efficient. The `quartic' coupling defined in Eq.~(\ref{quartic}) is only constant if $k=2$. In general, it scales as $\phi_0^{k-2}$. Nevertheless, for $k>4$ reheating from gravitational scattering is possible,
though very inefficient. For example, for $k=6$, $\trh \lesssim 1$ eV. As a result it is usually necessary to include a decay channel for the inflaton as in Eq.~(\ref{phidecaycoupling}).\footnote{Note that even including non-perturbative effects including preheating, does not lead to reheating in the absence of a decay channel for $k=2$ \cite{GKMOV}.}
For a sufficiently large coupling, $y$,
the radiation produced by decay will always dominate over that produced by scattering as computed above. 
In addition, the maximum temperature may be greater than the lower bound in Eq.~(\ref{Eq:tmax}). However, there is a critical value of $y$, such that at smaller couplings, the gravitational scattering process (\ref{Eq:rsigmah}) dominates at some point during
the reheating process. This gives us the reheating temperature below 
which the maximal temperature is fixed by (\ref{Eq:rhomaxbis}), and is independent
of additional couplings beyond gravity between the inflaton and the standard model sector. 
To determine the value of this critical coupling (and hence reheating temperature), it is useful to rewrite 
Eq.~(\ref{Eq:rhoR}) as 
\begin{eqnarray}
\rho^y_R&=& \frac{\sqrt{3}M_P^4\gamma_k^3 y^2 \Sigma^y_k}{8 \pi}
\left(\frac{\rhoe}{M_P^4}\right)^{\frac{k-1}{k}} \lambda^{-\frac2k}
\frac{k+2}{7-k} \nonumber \\
&& \times \left[
\left(\frac{\ae}{a}\right)^{\frac{6k-6}{k+2}} - \left(\frac{\ae}{a}\right)^4 \right]
\label{Eq:rhorsigma_sol2}
\end{eqnarray}
After 
some algebra, we found that the maximum of $\rho_R^y$ when evaluated at $a_{\rm max}$ given by Eq.~(\ref{Eq:amax}) is
\begin{eqnarray}
\rho_{\rm max}^y & = & \frac{y^2 \gamma_k^3 \sqrt{3}}{16 \pi}
\lambda^{\frac{-2}{k}}
M_P^4 \left(\frac{\rhoe}{M_P^4}\right)^{1-\frac{1}{k}}
\left(\frac{3k-3}{2k+4}\right)^{\frac{3k-3}{7-k}} \nonumber \\
& & \times \Sigma_k^y
\end{eqnarray}
where 
\beq
\Sigma_k^y = \sum_{n=1}^\infty n^3 |{\cal P}_n^k|^2 \, .
\label{Eq:sumyk}
\eeq
For $k=2$, the dominant mode is the first mode ($n=1$) which gives
\beq
\Sigma_2^y=1^3\times |{\cal P}_1^2|^2=\frac{1}{4}
\eeq
The critical value for $y$ such that the maximum radiation density and temperature are determined from the scattering of the inflaton condensate is given by $\rho_{\rm max}^y < \rho_{\rm max}^h$ which leads to
\bea
y^2 &\lesssim& N \frac{\lambda^{\frac{2}{k}}}{\gamma_k^2} \left(\frac{\rhoe}{M_P^4}\right) \frac{\Sigma^h_k}{\Sigma^y_k}
\nonumber
\\
&&\times
\left(\frac{k+2}{12k-6}\right)
\left(\frac{2k+4}{6k-3}\right)^{\frac{2k+4}{4k-7}}
\left(\frac{2k+4}{3k-3}\right)^{\frac{3k-3}{7-k}}
\label{ycrit}
\eea
which gives for $k=2$ and $N = 4$,
\beq
y\lesssim 0.4
\sqrt{\frac{\rhoe}{M_P^4}}
\simeq 
6.9 \times 10^{-6}\left(\frac{\rhoe}{10^{64} {\rm GeV}^4}\right)^{\frac{1}{2}}
\label{Eq:ymin}
\eeq
or
\beq
T_{\rm RH}\lesssim  3.0 \times 10^9 \left(\frac{\rhoe}{10^{64} {\rm GeV}^4}\right)^{1/2} \left(\frac{\lambda}{2.5 \times 10^{-11} }\right)^{1/4}  \,
\rm{GeV} \, ,
\label{maxtreh}
\eeq
where $\trh$ is defined by \cite{GKMO2}
\beq
\rho_\phi(\arh)=\alpha \trh^4=M_P^4\left(\frac{\sqrt{3}\gamma_k^3 y^2\Sigma_k^y\lambda^{-\frac{2}{k}}(k+2)}{8\pi(7-k)}\right)^k \, .
\label{rhoRH}
\eeq
Thus for all models with a reheat temperature due to decays, which is less than that given in Eq.~(\ref{maxtreh}), the maximum temperature during the reheat process is determined by scattering (mediated by gravity) and thus can not be ignored. 
Note also that for such small values of $y$, the kinetic effects due to the
effective mass induced by the coupling $y \phi \bar f f$ 
are non-existent, as shown in \cite{GKMO2}.

We show in Fig.~\ref{Fig:rhoR_num} the evolution of the energy densities of the inflaton (blue), the radiation produced by 
inflaton decays (orange dashed), the radiation produced by inflaton scattering mediated by gravity (green dashed), and the total radiation density (red) as function of the scaling 
parameter $a/a_{\rm end}$ for a Yukawa-like coupling $y=10^{-8}$ with $k=2$
and $\rhoe = 10^{64}~\rm{GeV^4}$. 
We clearly see that the beginning of the evolution of the radiation density
is dominated by the scattering of the inflaton via graviton exchange (orange line),  which determines the maximum temperature. 
For $k=2$, the radiation density from scattering falls as
$a^{-4}$ \cite{GKMO2}, whereas the density from decays falls more slowly as $a^{-3/2}$ so that eventually the latter begins to dominate the population of the thermal bath
when $a = a_{\rm int}$, until the reheating is complete when $\rho_\phi = \rho_R$ at $a = a_{\rm RH}$. 
For $a_{\rm int} \gg a_{\rm end}$, we can approximate the 
cross-over point from Eqs.~(\ref{Eq:rhorsigma_sol}) and (\ref{Eq:rhorsigma_sol2}) using the equality $\rho_R^y = \rho_R^h$. For sufficiently small $y$ and for $k=2$, we find
\beq
\frac{a_{\rm int}}{a_{\rm end}} \simeq \left(\frac{8 y^2 \Sigma^y_2}{5 N \Sigma^h_2} \frac{M_P^4}{\rhoe}  \right)^{-\frac25} \, ,
\label{aint}
\eeq
which gives $a_{\rm int} \simeq 430~a_{\rm end}$ in good agreement with the numerical solution for the parameter choices used in Fig.~\ref{Fig:rhoR_num}.
We stress that the maximum temperature attained 
$\tmax \simeq 10^{12}$ GeV is independent of any beyond the Standard Model physics, and is purely gravitational and can not be ignored when 
production rates are highly dependent on the ratio $\tmax / \trh$.

\begin{figure}[!ht]
\centering
\vskip .2in
\includegraphics[width=3.in]{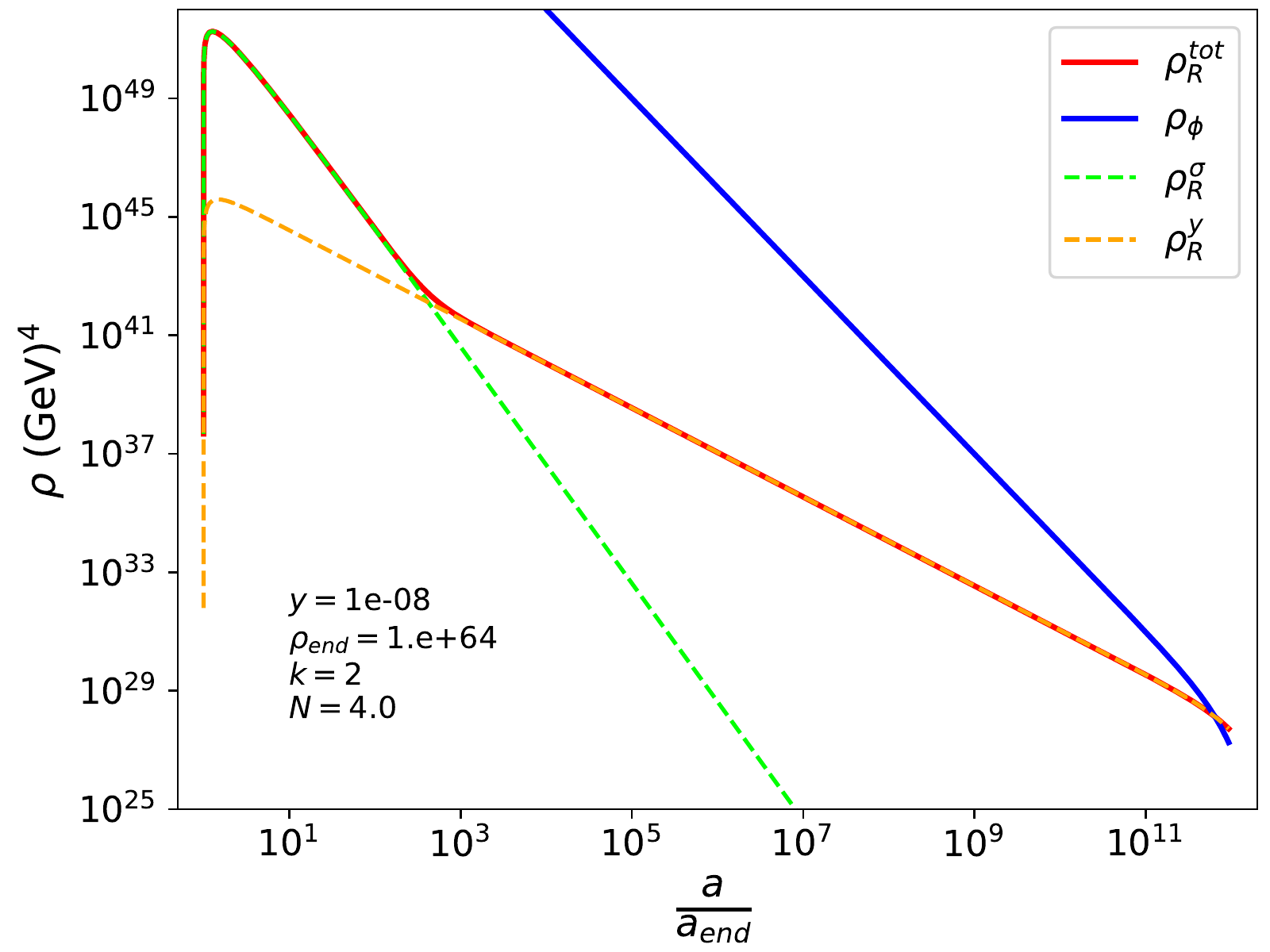}
\caption{\em \small Evolution of the radiation density (red) and inflaton density (blue) as a function of $a/a_{\rm end}$ for a Yukawa-like coupling $y=10^{-8}$ with $\rhoe = 10^{64}$ GeV$^4$ and $k=2$. This plot is obtained by solving numerically equations (\ref{Eq:diffrhophi}), (\ref{Eq:diffrhor}) and (\ref{Eq:rhorsigma}). The evolution of the radiation density produced from inflaton decays (orange-dashed) and scattering mediated by gravity (green-dashed) are also shown. }
\label{Fig:rhoR_num}
\end{figure}

We can finally apply our result to the dark matter production through a graviton exchange while the bath is also dominated by scattering of
$\phi$ through graviton exchange. For $T_{\rm RH} \lesssim 10^9$ GeV, the Boltzmann equation one needs to consider is

\beq
\frac{dY_X^h}{da}=\frac{\sqrt{3}M_P}{\sqrt{\rhoe}}a^2\left(\frac{a}{\ae}\right)^{\frac{3k}{k+2}}R_X^h(a)
\label{Eq:boltzmann5}
\eeq
with
\beq
R_X^h=\beta_X\frac{\rhomax^2}{\alpha^2M_P^4}\left(\frac{\amax}{a}\right)^8 \, .
\eeq
The result of the integration gives
\bea
Y_X^h(a_{\rm int}) &&= \frac{N^2 3\sqrt{3}M_P^3\beta_X \gamma_k^2(\Sigma_k^h)^2}{\alpha^2 65536\pi^2} \left(\frac{k+2}{8k -14}\right)^2
\nonumber
\\
 \times && \left(\frac{\rhoe}{M_P^4}\right)^{\frac{7k-4}{2k}} \ae^3 \left[ \left(\frac{k+2}{2k+10}\right) \left(1-\left(\frac{\ae}{a_{\rm int}}\right)^{\frac{2k+10}{k+2}}\right) \right.
\nonumber
\\
&&+ \left(\frac{k+2}{18k-18}\right)\left(1-\left(\frac{\ae}{a_{\rm int}}\right)^\frac{18k-18}{k+2}\right) \nonumber
\\
&&- \left. \left(\frac{k+2}{5k-2}\right)\left(1-\left(\frac{\ae}{a_{\rm int}}\right)^{\frac{10k-4}{k+2}}\right) \right]
\label{Eq:yield_bis}
\eea
where $a_{\rm int}$ corresponds to the value of the scale factor when the
radiation density produced by inflaton decays begins dominate over that produced by gravitational inflaton scattering (this only occurs if $y$ satisfies the bound in Eq.~(\ref{ycrit})). For $a > a_{\rm int}$, the
slope of the radiation energy density curve as a function of $a$ changes as seen in Fig.~\ref{Fig:rhoR_num} and any thermal contribution to the production of dark matter originates from inflaton decay.

 For sufficiently small $y$, $a_{\rm int} \gg a_{\rm end}$, and Eq.~(\ref{Eq:yield_bis}) can be simplified and we see that the dark matter yield does not depend on this intermediate scale factor, but only on $\ae$ and $\rhoe$. 
Thus for small $y$, we can also use Eq.~(\ref{Eq:yield_bis}) to evaluate the dark matter density at $a=a_{\rm RH}$,
\bea
n_X^h(a_{RH}) &&\simeq \frac{N^2\sqrt{3} \beta_X \gamma_k^2(\Sigma_k^h)^2 M_P^3}{196608\pi^2\alpha^2}\left(\frac{k+2}{8k-14}\right)^2
\\
&&\times \frac{(k+2)(4k-7)^2}{(k-1)(k+5)(5k-2)} \left(\frac{\rhoe}{M_P^4}\right)^{\frac{7k-4}{2k}}\left(\frac{\rhorh}{\rhoe}\right)^{\frac{k+2}{2k}}
\nonumber
\eea
and the relic abundance 
\bea
\Omega^h_X h^2 &&=1.6 \times 10^8 \frac{g_0}{g_{\rm RH}} \frac{m_X}{1~\rm{GeV}} \frac{\sqrt{3} \beta_X \gamma_k^2(\Sigma_k^h)^2 M_P^3}{196608\pi^2\alpha^2 \trh^3}\left(\frac{k+2}{8k-14}\right)^2
\nonumber
\\
&&\times \frac{(k+2)(4k-7)^2}{(k-1)(k+5)(5k-2)} \left(\frac{\rhoe}{M_P^4}\right)^{\frac{6k-6}{2k}}\left(\frac{\rhorh}{M_P^4}\right)^{\frac{k+2}{2k}}
\label{ohXh}
\eea
Because the radiation produced by gravitational scattering dominates near $a_{\rm max}$ only when $\trh$ satisfies Eq.~(\ref{maxtreh}), the relic density in 
Eq.~(\ref{ohXh}) is suppressed by $(\rho_{\rm RH}/M_P)^{(k+2)/2k}$, and it never dominates the {\it gravitational} production of dark matter given in Eq.~(\ref{22}), though it can lead to important effects when 
non-gravitational production modes with
a strong dependence on temperature are considered.

\section{Results}
\label{Sec:results}

In the results presented below, we choose
a class of inflation models, called T-models
given by Eq.~(\ref{Vatt}) which take the form of Eq.~(\ref{Eq:potmin}) when expanded about the origin. Given a specific potential, we can determined $\lambda$ from the normalization of the CMB quadrupole anisotropy and $\rho_{\rm end}$ from the condition $\epsilon = 1$, as discussed earlier. Setting $y=10^{-7}$, for $k=2$, we have $\lambda=2.5 \times 10^{-11}$  and 
$\rho_{\rm end}^{1/4} = 5.2 \times 10^{15}$ GeV\footnote{Different values of $y$ give differences (at most) of 20\% on $\lambda$, and 5\% on $\rhoe^{1/4}$.}. For $k=4$, $\lambda=3.3\times 10^{-12}$ and $\rho_{\rm end}^{1/4}=4.8\times 10^{15}\,{\rm GeV}$  whereas for $k=6$, 
$\lambda=4.6\times 10^{-13}$ and $\rhoe^{1/4}=4.6\times 10^{15}\,{\rm GeV}$. For more on the determination of these parameters, see \cite{GKMO2}. 

Given these (model-dependent) parameter values for $k = 2,4,6$, we list in Table~\ref{Tab:table}, the values for $\tmax$ which we obtain from $\rho^h_{\rm max}$ in Eq.~(\ref{Eq:rhomaxbis}); the maximum coupling $y$ from Eq.~(\ref{ycrit}) for which the gravitational produced radiation with temperature $\tmax$ dominates over that produced by decays; and the corresponding reheating temperature, ${\trh}_{\rm max}$ obtained when $y = y_{\rm max}$ using Eq.~(\ref{rhoRH}) for $\rho_{\rm RH}$. $\tmax \propto \lambda^{1/4k}$ depends weakly on the inflaton coupling, and thus varies little for different values of $k$.
The coupling $y_{\rm max}$ is independent of $\lambda$ and also varies little with k. However, the final reheat temperature (which is not a result of purely gravitational interactions) is very sensitive to $k$ as it scales as $y^{k/2}$ resulting in very small reheating temperatures when
$k = 4$ or 6 for the small values of $y$ considered.

\begin{table*}[!ht]
\begin{tabular}{|c|c|c|c|}
\hline 
  & $k=2$ & $k=4$ & $k=6$   
 \\ 
 \hline \hline
$\tmax $& $1.0\times 10^{12}$ GeV & $7.5\times 10^{11}$ GeV & $6.5\times 10^{11}$ GeV  
\\
\hline 
$y_{\rm max}$ & $1.8\times 10^{-6}$ & $1.4\times 10^{-6}$ & $1.1\times 10^{-6}$ 
\\
 \hline 
${\trh}_{\rm max}$ & $7.9\times 10^8~{\rm GeV}$ & 470 GeV  & $9.7\times10^{-4}$ GeV 
\\
\hline
\end{tabular}
\caption{Lower bound on $\tmax$ generated by the process $\phi \phi \rightarrow h_{\mu \nu} \rightarrow HH$ for different values of $k$. The radiation from this gravitational scattering dominates when $y < y_{\rm max}$ and we also list the corresponding reheating temperature $\trh$ when $y = y_{\rm max}$. }
\label{Tab:table}
\end{table*}

\begin{figure}[!ht]
\centering
\vskip .2in
\includegraphics[width=3.in]{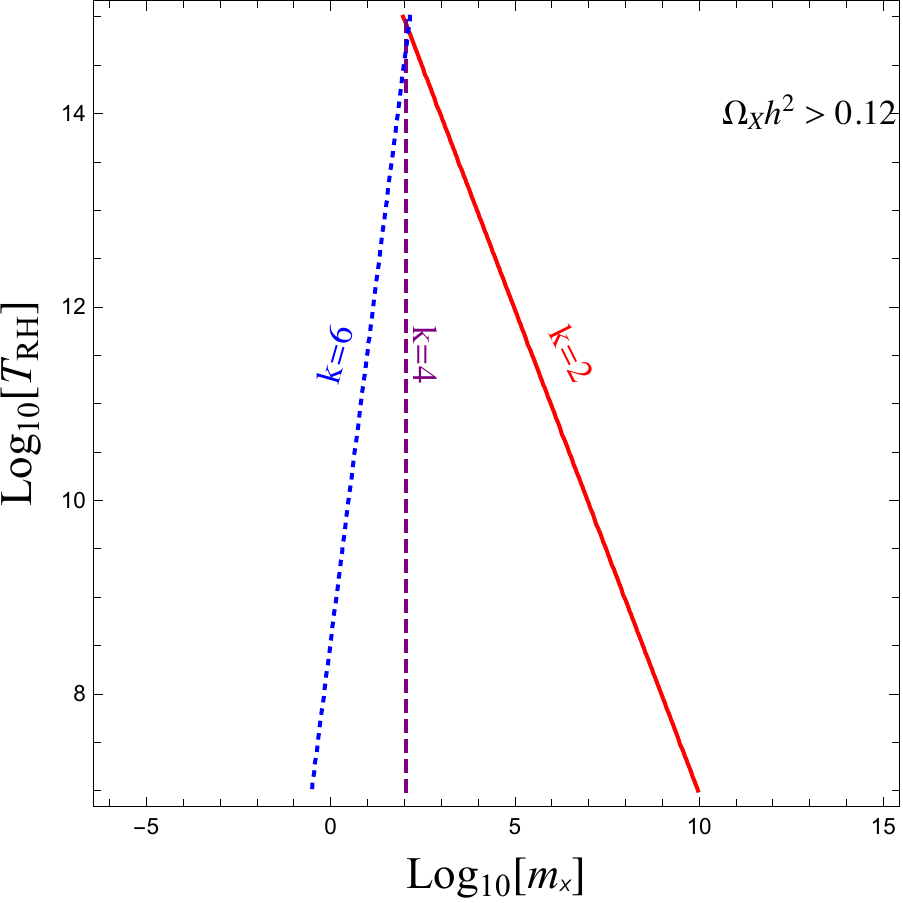}
\caption{\em \small Points respecting Planck constraint $\Omega h^2=0.12$ in the case of a scalar dark matter, in the plane 
($m_X$, $\trh$) for  different values of $k$.}
\label{Fig:omega00}
\end{figure}

\begin{figure}[!ht]
\centering
\vskip .2in
\includegraphics[width=3.in]{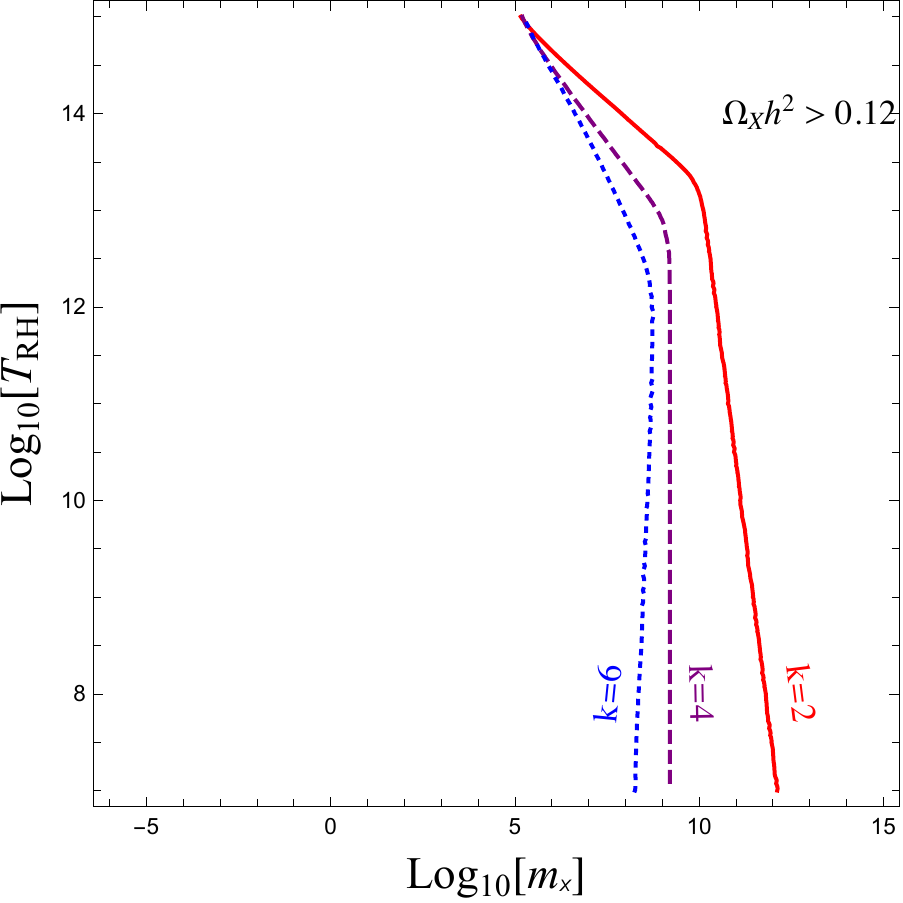}
\caption{\em \small Points respecting Planck constraint $\Omega h^2=0.12$ in the case of a fermionic dark matter, in the plane 
($m_X$, $\trh$) for different values of $k$.}
\label{Fig:omegahalf0}
\end{figure}

We show in Figs.(\ref{Fig:omega00}) and (\ref{Fig:omegahalf0}) (for scalar and fermionic dark matter respectively) the region in the parameter space defined by the
($m_X$, $\trh$) plane for which we are able to obtain a relic abundance consistent with the Planck CMB determination of the cold dark matter density, 
$\Omega_X h^2=0.12$ \cite{Planck}. 
We combine the dark matter density originating from thermal production 
as given in Eq.~(\ref{22}) with that from scattering of the condensate
to scalars given in Eq.~(\ref{Eq:omega0}) or fermions in Eq.~(\ref{Eq:omegaphihalf}).

For scalar dark matter, scattering in the condensate dominates the production of dark matter 
and we see from Eq.~(\ref{Eq:omega0}) that an isodensity contour should obey
 a simple power law, corresponding to $m_X\propto (\trh)^{\frac{4}{k}-1}$.
Indeed, thermal production is not an efficient mechanism for the scalar dark matter, and the 
unique mechanism which populates the dark matter density is inflaton scattering (barring any beyond the Standard Model contribution).
To better understand this, we can compute the ratio of the rates when $a = a_{\rm max}$, where the thermal production is maximum. Comparing the rates in Eqs.~(\ref{Eq:ratephi0}) and (\ref{rateXT})
\bea
&&
\frac{R^{\phi^k}_0(a_{\rm max})}{R^T_0(a_{\rm max})}
=\frac{\alpha^2\Sigma^k_0}{8 \pi \beta_0}\left(\frac{3k-3}{2k+4}\right)^{\frac{6}{7-k}}
\left(\frac{\rhoe}{\rho_{\rm RH}}\right)^{\frac{2}{k}}
\\
&&=g_{\rm max}^2\frac{5760 \Sigma^k_0}{3997}
\left(\frac{3k-3}{2k+4}\right)^{\frac{6}{7-k}}
\left(\frac{\rhoe}{\rho_{\rm RH}}\right)^{\frac{2}{k}}
\gg 1 \, ,
\nonumber
\eea
where $g_{\rm max} = 427/4$ is the number of degrees of freedom at $a_{\rm max}$
in the Standard Model. Since $\rhoe \gg \rhorh$, we clearly see
that the ratio is much greater than one. This implies that the gravitational production is always dominated by the scattering of the inflaton zero modes.

Restricting our attention to Eq.~(\ref{Eq:omega0}) for the production of dark matter scalars, we see that for $k=4$, something interesting happens. The relic abundance is independent of the reheating process, and depends
only on the energy density at the end of inflation. This comes from the fact that for increasing values of $k$, the production of dark matter is less efficient, and from Eq.~(\ref{n0phi}), we see that $n_0^\phi\propto \trh^3$ for $k=4$. Dilution effects thus render the present abundance independent of $\trh$ and there is a unique universal limit of $ m_X\lesssim 120$ GeV
for scalar dark matter and $\lesssim 1.7 \times 10^9$ GeV for fermionic dark matter (when inflaton scattering dominates). 
For $k>4$, the slope of $\trh$ vs $m_X$ changes sign, 
and the required reheating temperature grows with the dark matter mass.
In this case,  even sub-GeV dark matter candidates are allowed  for low reheating temperatures, whereas for 
 $k=2$ and $k=4$ the production process is too weak to produce MeV
dark matter in sufficient quantities to account for the cold dark matter density as determined by Planck \cite{Planck}.

The ($m_X$, $\trh$) plane for fermionic dark matter is shown in Fig.~\ref{Fig:omegahalf0}. In this case both the scattering from a condensate and thermal gravitational contributions must be considered. Notice that there is a change in slope between the required reheating temperature and dark matter mass. For higher masses, the scattering from the condensate dominates as in the case of scalar dark matter and we require $m_X \propto \trh^{\frac{(k-4)}{3k}}$ as can be seen from Eq.~(\ref{Eq:omegaphihalf}). However, at lower masses, because of the mass suppression in the rate in Eq.~(\ref{eq:rateferm}) and hence the abundance of dark matter in Eq.(\ref{Eq:nhalf}), there is a region where the thermal production dominates over $\phi-\phi$ scattering. In this case, $m_X\propto \trh^{-3}, \trh^{-2}, \trh^{-1}$,
for $k=2, 4$ and 6 respectively, as can be seen from Eq.~(\ref{22}). The origin of this suppression is simply a helicity argument;  the scattering of two scalars generates rates
where a spin-flip is required making it proportional to
the mass of the fermion in the final state. Thus the rate vanishes for a massless fermion. This is not the case for thermal production, because Standard Model particles in the thermal bath
are relativistic and then can still produce fermionic dark matter through scattering without being affected by a helicity suppression. 
To be more quantitative, we again compare the production rates in Eqs.~(\ref{eq:rateferm}) and (\ref{rateXT}) at $a = a_{\rm max}$
\bea
&&
\frac{R^{\phi^k}_{1/2}(a_{\rm max})}{R^T_{1/2}(a_{\rm max})}
=\frac{\alpha^2\Sigma^k_{1/2}}{ 2 \pi \beta_{1/2}}\frac{m_X^2}{m_\phi^2}\left(\frac{3k-3}{2k+4}\right)^{\frac{6}{7-k}}
\left(\frac{\rhoe}{\rho_{\rm RH}}\right)^{\frac{2}{k}}
\nonumber
\\
&&=g_{\rm RH}^2\frac{11520~\Sigma^k_{1/2}}{11351}
\frac{m_X^2}{m_\phi^2}
\left(\frac{3k-3}{2k+4}\right)^{\frac{6}{7-k}}
\left(\frac{\rhoe}{\rho_{\rm RH}}\right)^{\frac{2}{k}} \, .
\eea
In contrast to the scalar case, we see that there exists a value of $m_X \lesssim (.13,.050,.036) (\rho_{\rm RH} / \rhoe)^{1/k} m_\phi$, for $k=2, 4$ and 6 respectively for which the relic abundance is dominated by the thermal
production.

\section{Conclusions}
\label{Sec:conclusion}

We have considered the production of matter and radiation interacting only gravitationally with the inflaton through the exchange 
of a graviton $h_{\mu \nu}$. We compared the production of dark matter from inflaton scattering
and from the thermal bath (mediated only by gravity). The former tends to dominate the production in a large part of the parameter space. However, we noticed a notable difference in the case of fermionic dark matter, because the production through $\phi \phi$ scattering
is suppressed by a mass flip proportional to the dark matter mass $m^2_X$. We have also seen that it is possible to produce radiation from inflaton scattering in the condensate during the earlier stages of reheating. As a result, we have 
derived a {\it lower bound} on the maximal temperature is expected
from $\phi \phi \rightarrow h_{\mu \nu} \rightarrow HH$ of
the order of $10^{12}$ GeV for a typical chaotic or $\alpha-$attractor scenario. This
lower gravitational bound becomes the effective maximal
temperature for $\trh \lesssim 10^9$ GeV (for $k=2$.
 As a conclusion, gravitational effects gives lower bounds on maximal temperature and relic abundance that cannot be neglected and should be considered as the minimal ingredients to add to any non-minimal extension of the Standard Model. 
During the final phase of our work, a paper conducting a similar analysis appeared
\cite{Haque:2021mab}. The results obtained are largely in agreement with our own.

\noindent {\bf Acknowledgements. }  The authors thank Marcos Garc{\' i}a and Kunio Kaneta for useful discussions. This work was made possible by with the support of the Institut Pascal at 
Université Paris-Saclay during the 
Paris-Saclay Astroparticle Symposium 2021, with the support of the P2IO Laboratory of Excellence (program “Investissements d’avenir” ANR-11-IDEX-0003-01 Paris-Saclay and ANR-10-LABX-0038), the P2I axis of the Graduate School Physics of Université Paris-Saclay, as well as IJCLab, CEA, IPhT, APPEC, the IN2P3 master projet UCMN and
ANR-11-IDEX-0003-01 Paris-Saclay and ANR-10-LABX-0038. This project has received support from the European Union’s Horizon 2020 research and innovation programme under the Marie Sk$\lslash$odowska-Curie grant agreement No 860881-HIDDeN and the CNRS PICS MicroDark. The work of K.A.O.~was supported in part by DOE grant DE-SC0011842  at the University of
Minnesota.

\section*{Appendix}
\appendix
\renewcommand{\thesubsection}{\Alph{subsection}}
\subsection{THERMAL PRODUCTION}
\label{sec:appendixA}
In this appendix, we describe our calculation of the production 
rate for scalar and fermionic dark matter, and include the amplitude squared for the relevant processes. If we ignore the masses of Standard Model particles, 
the rate $R(T)$ for the processes ${\rm{SM}}+{\rm{SM}} \rightarrow {\rm{DM}}_j+{\rm{DM}}_j$ can be computed from 
\begin{widetext}
\begin{equation}
     R_j^T=\sum_{i=0,1/2,1} N_i R_{ij}=\sum_{i=0,1/2,1}
     \frac{N_i}{1024 \pi^6}\int f_i(E_1) f_i(E_2) E_1 \diff E_1 E_2 \diff E_2 \diff \cos \theta_{12}\int |{\cal M}^{ij}|^2 \diff \Omega_{13}  =   4R_{0j}+45R_{\frac{1}{2}j}+12R_{1j}  \, ,   
    \label{Eq:totalrate}
\end{equation}
\end{widetext}
where $N_i$ denotes the number of each SM species of spin $i$: $N_0=4$ for 1 complex Higgs doublet, $N_1=12$ for 8 gluons and 4 electroweak bosons, and $N_{1/2}=45$ for 6 (anti)quarks with 3 colors, 3 (anti)charged leptons and 3 neutrinos, c.f., Eq.~(\ref{Eq:ampscat}).
The infinitesimal solid angle is defined as
\begin{equation}
    \diff \Omega_{13}=2 \pi \diff \cos \theta_{13} \, ,
\end{equation}
with $\theta_{13}$ and $\theta_{12}$ being the angle formed by momenta ${\bf{p}}_{1,3}$ and ${\bf{p}}_{1,2}$, respectively. In the massless limit, one can express the amplitude squared in terms of Mandelstam variables, $s$ and $t$, which are related to the angles $\theta_{13}$ and $\theta_{12}$ by the expressions
\begin{align}
    t\,= & \, \dfrac{s}{2}\left( \cos \theta_{13}-1 \right) \, ,  \\  s\, = & \,2E_1E_2 \left(1-\cos \theta_{12} \right) \, .
\end{align}
The amplitudes and rates for scalar and fermionic dark matter are given in the following subsections. 

\subsection*{Scalar dark matter}

We note that we include the symmetry factors of the initial and final states in the squared amplitudes, and indicate it with an overbar:
\bea
&&
|\overline{{\cal M}}^{00}|^2= \frac{1}{4 M_P^4} \frac{t^2(s+t)^2}{s^2} \, ,
\\
&&
|\overline{{\cal M}}^{\frac{1}{2} 0}|^2 = \frac{1}{4 M_P^4} \frac{(-t(s+t))(s+2t)^2}{s^2} \, ,
\\
&&
|\overline{{\cal M}}^{1 0}|^2 = \frac{1}{2 M_P^4} \frac{t^2(s+t)^2}{s^2} \, .
\eea
Using these amplitudes in Eq. (\ref{Eq:totalrate}), we obtain \cite{Bernal:2018qlk}
\bea
&&
R^T_0=
\frac{3997\pi^3}{41472000} \frac{T^8}{M^4_P} 
\equiv \beta_0 \frac{T^8}{M^4_P} \, .
\label{R0a}
\eea

\subsection*{Fermionic dark matter}

The corresponding amplitudes for fermionic dark matter are given by:
\begin{equation}
|\overline{{\cal M}}^{0\frac{1}{2}}|^2 = \frac{(-t(s+t))(s+2t)^2}{4 M_P^4 s^2} \, , 
\end{equation}
\begin{equation}
|\overline{{\cal M}}^{\frac{1}{2}\frac{1}{2}}|^2 = \frac{s^4+10s^3t+42s^2t^2+64st^3+32t^4}{8 M_P^4 s^2} \, ,
\end{equation}
\begin{equation}
|\overline{{\cal M}}^{1 \frac{1}{2}}|^2 = \frac{(-t(s+t))(s^2+2t(s+t))}{M_P^4 s^2} \, ,
\end{equation}
which leads to the following rate \cite{Bernal:2018qlk}
\begin{equation}\begin{split}
& R^T_{\frac{1}{2}} = \frac{11351 \pi ^3}{20736000 } \frac{T^8}{M_P^4} 
\equiv \beta_{\frac{1}{2}} \frac{T^8}{M^4_P} \, .
\end{split}\end{equation}

\subsection{INFLATON CONDENSATE SCATTERING}
\label{sec:appendixB}

In this appendix, we describe our calculation of the particle production rate of dark matter from the scattering of the inflaton condensate. If we consider the gravitational scattering process $\phi(p_1) + \phi({p_2}) \rightarrow X^i(p_3) + X^i(p_4)$, with $i = 0, 1/2$, illustrated by the Feynman diagram in Fig.~\ref{Fig:feynman}, the Boltzmann equation for the number density of produced dark matter particles is given by \cite{Nurmi:2015ema,GKMO2}
\begin{equation}
    \frac{dn_{X}}{dt} + 3H n_{X} \; = \; R_i^{\phi^k} \, ,
\end{equation}
where the rate is given by
\bea
&R_i^{\phi^k} \; \equiv \; g_X \int d \Psi_{1, 2, 3, 4} (2\pi)^4 \delta^{(4)} \left(p_1 + p_2 - p_3 - p_4 \right)  \nonumber \\
&~\times \Big[|\mathcal{M}|^2_{12 \rightarrow {34}} f_1 f_2 (1 \pm f_3) \left(1 \pm f_4 \right)  - \left(34 \leftrightarrow 12 \right) \Big] \, 
\eea
where $d \Psi_{1, 2, 3, 4} = \Pi_{i = 1}^4 d^3 {\bf{p}}_i/\left((2\pi^3)2p^0_i \right)$ denotes the phase space distribution of particles $1, 2, 3$ and $4$, $\mathcal{M}$ is the transition amplitude, $f_i$ is the phase space density of species $i$, and $g_X$ denotes the number of produced dark matter particles. If we ignore the Bose enhancement and Pauli blocking effects, the above rate can be approximated as
\bea
    R_i^{\phi^k}  &=& g_X \int  \frac{d^{3} {\bf{p}}_{3}}{(2 \pi)^{3} 2 p_{3}^{0}} \frac{d^{3} {\bf{p}}_{4}}{(2 \pi)^{3} 2 p_{4}^{0}} \nonumber \\
    &\times& (2 \pi)^{4} \delta^{(4)}\left(p_{1} + p_2 -p_{3}-p_{4}\right)|\mathcal{M}|^2_{12 \rightarrow {34}} \, .
\eea
For the inflaton condensate we can use the transition amplitude $\mathcal{M}_n$ for each oscillating field mode of $\phi$. In this case, the four-momentum of the $n$-th oscillation mode is given by $p_1 + p_2 = p_n = \sqrt{s} = (E_n, \bf{0})$ with $E_n$ the energy of the $n$-th oscillation mode. Since the transition amplitude $\mathcal{M}_n$ of the $n$-th oscillation does not depend on the final particle momenta ${\bf{p}}_{3,4}$, we can approximate the rate as
\begin{equation}
    \label{eq:apprate1}
    R_i^{\phi^k} \; = \; \frac{g_X}{l!} \frac{1}{8\pi} \sum_{n = 1}^{\infty} |\mathcal{M}_n|^2 \sqrt{1 - \frac{4 m_{X}^2}{s}} 
\end{equation}
where $l$ is associated with the number of identical particles in the final state.

For the production of scalar dark matter, we find that the scattering amplitude squared is given by
\begin{equation}
    |{\mathcal{M}}^{0 \phi^k}_n|^2 = \frac{\rho_{\phi}^2}{M_P^4} \left[1 + \frac{2m_X^2}{s} \right]^2 |({\mathcal{P}^k})_n|^2 \, ,
\end{equation}
where $s = E_n^2 = n^2 \omega^2$, and we used $\rho_{\phi} = \frac{\lambda \phi^{k}}{M_P^{k-4}}$. We find that the inflaton scattering rate is given by
\begin{equation}
R^{\phi^k}_0=\frac{2 \times \rho_{\phi}^2}{16\pi M_P^4} \sum_{n=1}^{\infty} \left[1 + \frac{2m_X^2}{E_n^2} \right]^2 |({\mathcal{P}^k})_n|^2 \langle \beta_{n}\left(m_{X} \, , m_{X}\right) \rangle \, ,
\end{equation}
where 
\begin{equation}
\beta_{n}\left(m_{A}, m_{B}\right) \equiv \sqrt{\left(1-\frac{\left(m_{A}+m_{B}\right)^{2}}{E_{n}^{2}}\right)\left(1-\frac{\left(m_{A}-m_{B}\right)^{2}}{E_{n}^{2}}\right)} \, ,
\end{equation}
and we used $g_X = 2$. For the case $k = 2$, we find that the rate is given by Eq.~(\ref{eq:ratescalar2}).

Similarly, for fermionic dark matter we find that the scattering amplitude squared is,
\begin{equation}
    |{\mathcal{M}}^{1/2 \, \phi^k}_n|^2 = \frac{2\rho_{\phi}^2}{M_P^4} \frac{m_{X}^2}{s}\left[1 - \frac{4m_X^2}{s} \right] |({\mathcal{P}^k})_n|^2 \, ,
\end{equation}
and the rate is given by Eq.~(\ref{eq:ratefermion2}).

The rates as defined in the text depend on various summations over the Fourier modes of the periodicity function $\mathcal{P}(t)$. In Table~\ref{Tab:tableSigma}, the numerical values of these quantities are given for $k=2, 4, 6$. Values are given in the limit of vanishing dark matter mass.

\begin{table*}[!ht]
\begin{tabular}{|c|c|c|c|}
\hline 
  & $k=2$ & $k=4$ & $k=6$   
 \\ 
 \hline \hline
$\Sigma_0^k $ (Eq.~(\ref{Sigma0k})) & $\frac{1}{16}$ & 0.063  &  0.056
\\
\hline 
$\Sigma_{1/2}^k$ (Eq.~(\ref{eq:ratefermion2}))  & $\frac{1}{64}$ & 0.061 & 0.101
\\
 \hline 
$\Sigma_k^h$ (Eq.~(\ref{Eq:sumhk})) & $\frac18$ &  0.126 & 0.124
\\
\hline 
$\Sigma_k^y$ (Eq.~(\ref{Eq:sumyk})) & $\frac14$ & 0.241 & 0.244
\\
\hline
\end{tabular}
\caption{Numerical values of the various summations of the Fourier modes of the periodicity functions used in the text.
The dark matter mass has been neglected in producing the numerical values.
}
\label{Tab:tableSigma}
\end{table*}

\end{document}